\documentstyle[emulateapj]{article}

\newcommand{\msun}{$M/M_{\odot}\,$}
 
\newcommand{\vhbb}{$\Delta V_{\rm HB}^{\rm Bump}\,$} 

\begin{document}

\title{Comparison between predicted and empirical \vhbb 
in Galactic Globular Clusters}

\author{M. Zoccali\altaffilmark{1}, S. Cassisi\altaffilmark{2}, 
G. Piotto\altaffilmark{1}, G. Bono\altaffilmark{3}, M. Salaris\altaffilmark{4}}

\lefthead{Zoccali et al.}
\righthead{\vhbb in Galactic Globular Clusters} 

\altaffiltext{1}{Dipartimento di Astronomia, Universit\`a di Padova, Vicolo 
dell'Osservatorio 5, 35122 Padova, Italy; manu@pd.astro.it~~~piotto@pd.astro.it}
\altaffiltext{2}{Osservatorio Astronomico di Collurania, Via M. Maggini,
64100 Teramo, cassisi@astrte.te.astro.it} 
\altaffiltext{3}{Osservatorio Astronomico di Roma, Via Frascati 33, 
00040 Monteporzio Catone, Italy; bono@coma.mporzio.astro.it} 
\altaffiltext{4}{Astrophysics Research Institute, Liverpool John Moores 
University, Twelve Quays House, Egerton Wharf, Birkenhead L41 1LD, UK; 
ms@staru1.livjm.ac.uk}

\begin{abstract}

We present observational estimates of \vhbb in a sample of 28  
Galactic Globular Clusters (GGCs) observed by HST. 
The photometric accuracy and the sizable number of stars 
measured in each cluster allowed us to single out the RGB Bump both 
in metal-poor and in metal-rich GGCs.  

Empirical values are compared with homogeneous theoretical predictions 
which account for both H and He burning phases over a wide range of 
metal abundances ($0.0001 < Z < 0.02$).   
We found that, within current observational uncertainties on both 
iron and $\alpha$-element abundances, theory and observations are 
in very good agreement, provided that the metallicity scale by Carretta 
\& Gratton (1997) as extended by Cohen et al. (1999) is adopted. 
Interestingly enough, we also found that both theoretical and observed 
values show a change in the slope of the \vhbb-[M/H] relation toward 
higher metal contents. 
\end{abstract}



\section{Introduction}

The RGB Bump is an intrinsic feature of the RGB Luminosity Function 
(LF) of GGCs. It appears as a peak in the differential LF or as a 
change in the slope of the cumulative LF. The presence of the Bump is 
due to the fact that during the RGB evolution the H-burning shell
crosses the chemical discontinuity left over by the convective envelope 
soon after the first dredge-up phase.  
Since its first detection in 47 Tuc (King, Da Costa \& Demarque
1985), it became the crossroad of several theoretical and observational
investigations (Alves \& Sarajedini 1999).  
The detection of the RGB Bump was mainly hampered by
the size of the available samples of RGB stars. This is particularly
true for the most metal-poor clusters, where the Bump moves toward 
brighter magnitudes and therefore less populated RGB regions.
Only recently was this feature firmly detected in a large set of both 
Galactic (FP; Brocato et al. 1996) and extragalactic stellar systems 
(Desidera 1999).

The analysis by FP showed that the \vhbb\footnote{The
parameter \vhbb is commonly defined as the difference in magnitude 
between the Red Giant Branch (RGB) Bump and the Horizontal Branch 
(HB) stars located within the RR Lyrae instability strip (see for 
a detailed discussion Fusi Pecci et al. 1990, hereinafter FP, and 
Cassisi \& Salaris 1997, hereinafter CS).}
values predicted by theory 
were 0.4 mag brighter than the empirical estimates
in a sample of 11 GGCs.  This observable, which does not depend on 
the distance modulus, on the reddening, or on the calibration of
photometric data, is a key parameter for assessing both the accuracy
and the plausibility of the physical assumptions adopted for
constructing the evolutionary models. As a consequence the discrepancy
found by FP needs to be understood.
In order to explain this mismatch between theory and observations
Alongi et al. (1991) suggested the inclusion of convective
undershooting at the boundary between the convective envelope and the
thin H-burning shell, whereas Straniero, Chieffi \& Salaris (1992)
called attention on the role that the global metallicity could play on
the estimate of this observable. The latter hypothesis was confirmed
by CS who found that standard models agree quite well
with the observed \vhbb values in 8 GGCs for which high resolution
spectroscopic determinations of both [Fe/H] and [$\alpha$/Fe] were
available. The significance of this result was limited by the small
number of clusters taken into account and by the small metallicity
range that they cover.

The main aims of this investigation are to provide new homogeneous 
measurements of both the Bump position and the corresponding \vhbb value 
in a large sample of GGCs covering a wide metallicity range
($-2.1 \le $ [Fe/H] $ \le  -0.2$); and also to compare the new 
measurements with the predictions of standard evolutionary models.
At present, the most homogeneous photometric sample comes from HST. 
Two of us (GP and MZ) are involved in a program aimed at
collecting F439W and F555W WFPC2 images for all the GGC with
($m-M$)$_B<18.0$ not yet observed with HST. Ten of these
clusters have been observed during Cycle 6 (GO6095) and nine more
during the ongoing Cycle 7 (GP7470). Similar observations for nine
additional clusters are available on the HST archive.  

In \S 2 we present this sample of 28 GGCs, and discuss the data reduction
strategy  as well as the approach adopted for estimating both the 
zero age horizontal branch (ZAHB) and the RGB Bump visual magnitudes. 
Section 3 deals with the comparison between theory and observations, 
while in \S 4 we briefly summarize the main results.   

\section{Data Selection and Reduction}

The center of all the clusters were observed in the HST B (F439W)
and V (F555W) bands with the WFPC2. Both very short (a few seconds)
and long (a few hundreds of seconds) exposures are available for both
filters allowing us to map the evolved regions of the color-magnitude
diagram (CMD). The pre-processing, photometric reduction, and
calibration to a standard B and V system of the data for each cluster
were carried out following the same procedure as described in details
in Piotto et al.\ (1999a) for a subset of images of the GO7470.  The
photometry for all the clusters extends from the tip of the RGB down
to about 2 magnitudes below the turnoff, and include a number of stars
ranging from $\sim 5000$ for the loosest cluster to $\sim 30000$ for
the most concentrated ones. A detailed description of the CMDs will be
presented in Piotto et al. (1999b).
The large sample of stars, coupled with the high photometric accuracy,
allows us to identify of the RGB Bump even in the most metal-poor clusters. 


Figure 1 shows an example of the Bump identification in NGC5824, a 
metal-poor cluster in our database.
The upper left panel shows the RGB differential LF. We constructed a
histogram with a fixed bin size of 0.15 magnitude ({\it dotted line})
as well as the multibin histogram ({\it solid line}) described in
Piotto et al. (1999c). The Bump is marked by a vertical arrow.  
As a further check of the accuracy in the Bump identification, the
lower left panel shows the RGB cumulative LF. The Bump can be easily
located since it is marked by a change in the slope.  The right panel
shows the CMD of NGC5824 and the arrow marks the position of the Bump
along the RGB.  

The determination of the ZAHB magnitude ($V_{\rm ZAHB}$)
is a thorny problem, in particular for those clusters which show blue
HBs. A standard method to determine the ZAHB magnitude for both
intermediate and metal-poor clusters is to adopt the mean magnitude of
the cluster RR Lyrae stars. Unfortunately, we could not adopt this
method directly, since our photometry covers a very short time interval, and
therefore the RR Lyrae were always measured at random pulsation phases
(Piotto et al. 1999a).

In order to overcome this problem we have undertaken a different
approach.  For metal-poor and intermediate metallicity clusters we
selected from the literature three clusters which offer accurate
photometry, a large number of RR Lyrae stars, and a well defined blue
HB tail.  The template clusters are the following: NGC1851
(Walker 1998) for clusters in the metallicity range
$-1.5<$[Fe/H]$<-1.0$, NGC5272 (M3, Buonanno et al. 1994) for clusters
in the range $-1.7<$[Fe/H]$<-1.5$, and NGC4590 (M68, Walker 1994) for
more metal-poor clusters. The CMDs of the template clusters were
artificially shifted in color and in magnitude in order to match both
the RGB and the blue HB tail of each cluster. The mean RR Lyrae magnitude 
of the template cluster was scaled according to the magnitude shift 
adopted to overlap the stellar distributions
along the HB.  The $V_{\rm ZAHB}$ magnitudes were estimated by
using the relation between the mean RR Lyrae magnitude and the
$V_{\rm ZAHB}$ magnitude suggested by CS. It is worth noting that this 
method is totally independent from any zero point difference between the 
template photometry and ours.

For the most metal-rich clusters ([Fe/H]$>-1$) with well populated red
HBs but no RR Lyraes, first we estimated the photometric error
$\sigma^2_V=\sigma^2_{(B-V)}/2$, where $\sigma^2_{(B-V)}$ is the
standard deviation of the color distribution of the RGB stars at the
level of the HB. Then we fixed $V_{\rm ZAHB}$ at $3 \sigma_V$ magnitudes
above the lower envelope of HB stellar distribution. However, in order
to provide a consistent determination of this parameter over the whole
GGCs sample, the $V_{\rm ZAHB}$ at the level of RR Lyrae instability strip
was evaluated according to the method suggested by Fullton et al. (1995).
Figure 1 shows the fit of the same metal-poor
cluster NGC5824, (full dots) with the template cluster
M68 (crosses). A vertical and a horizontal shift of 2.83 and 0.08 
magnitudes were applied to M68 for matching both the RGB and
the HB.  The M68 RR Lyrae stars are plotted as open squares.
Taking into account the uncertainties in the fit, we estimated 
an error in $V_{\rm ZAHB}$ of the order of 0.1 mag. 

Table 1 summarizes the cluster observables: column (1) gives 
the NGC and the Messier numbers; column (2) lists the visual magnitudes 
of the RGB Bump and the photometric error;  
column (3) gives the mean RR Lyrae visual magnitudes estimated according 
to the method previously described;  
column (4) lists the ZAHB visual magnitude at the lower envelope of red HB 
stars in metal-rich clusters; column (5) gives the cluster metallicities 
according to the Carretta \& Gratton (1997, hereinafter CG) 
scale\footnote{For those clusters whose metallicity was not provided by 
CG, the cluster metallicities collected by Harris (1996) were transformed 
into the CG scale by adopting the new calibration provided by Cohen 
et al. (1999, hereinafter CGBC). In contrast with CG the new scale applies 
also to metal-rich clusters ($-2.12 \le$ [Fe/H] $ \le -0.3$).}.

\section{Comparison Between Theory and Observations}

Figure 2 shows the comparison between theory and observations in the 
[M/H]-\vhbb plane. The global metallicities were estimated
by adopting a mean $\alpha$ enhancement equal to 0.3 for clusters with 
[Fe/H] $ < -1.0$ and to 0.20 for more metal-rich clusters. The former 
value was suggested by Carney (1996, hereinafter C96), while 
the latter, due to the paucity of data available in the literature, 
is a mean between the estimates collected by C96 and by Salaris \& 
Cassisi (1996, hereinafter SC).  
The observed \vhbb values have been plotted in the top panel according to 
the CG metallicity scale. 
The \vhbb error bars have been calculated by quadratically combining
the errors on $V_{\rm ZAHB}$ and on $V_{\rm Bump}$.  The global metallicity
error bars are a lower limit of the uncertainties affecting both [Fe/H] and 
[$\alpha$/Fe] measurements (see C96 and 
Rutledge, Hesser, \& Stetson 1998, hereinafter RHS). 


The theoretical predictions plotted in the top panel were estimated by
adopting progenitor masses ranging from \msun=0.8 to 1.0 and a wide
range of global metallicities ($-2.3 \le$ [M/H]$ \le 0.0$).  The initial
helium contents adopted in constructing evolutionary models are the
following: $Y=0.23$ for [M/H]$\le -0.5$, $Y=0.255$ for [M/H]$=-0.25$,
and $Y=0.289$ for [M/H]$=0.0$.  Metal-poor \vhbb theoretical estimates 
up to [M/H]$\approx -0.5$ already been presented in CS, 
whereas more metal-rich ones were specifically computed, 
to extend the predictions to clusters more metal-rich than 47 Tuc.
Basic assumptions on the input physics adopted for constructing
evolutionary models were extensively described in CS and 
Bono et al. (1997 and references therein) and therefore they are
not discussed here. The reader interested in a detailed
discussion on the dependence of \vhbb on mixing-length,
He content, and element diffusion is referred to CS and to 
Cassisi, Degl'Innocenti, \& Salaris (1997, hereinafter CDS).  
Bolometric magnitudes were transformed into V magnitudes by 
adopting bolometric corrections provided by Castelli, Gratton \& 
Kurucz (1997).

In order to account for the $V_{\rm Bump}$ dependence on cluster age
as suggested by CS and more recently by Alves \& Sarajedini (1999)
in a detailed investigation on HST data of 8 SMC clusters, we plotted
the theoretical predictions for three different ages: 12 (short-dash line),
14 (solid line), and 16 (long-dash line) Gyr. In the last few years a
large number of theoretical and observational investigations have been
devoted to the absolute and the relative ages of GGCs as well as to the
errors affecting such parameters. In fact, they depend on the physical
assumptions and on the input physics (Cassisi et al. 1998; Vandenberg,
Stetson \& Bolte 1996) adopted for constructing evolutionary models.
As plausible assumptions we adopted an average cluster age of 14 Gyr 
(Vandenberg 1999) and an average uncertainty of $\pm2$ Gyr. 
The change in the slope toward higher metal contents shown by theoretical 
predictions is due to the fact that at fixed age an increase in the evolving 
mass causes a smoother increase in the core-mass luminosity relation, 
and in turn in the $V_{\rm Bump}$ magnitude. 
In fact, for metallicities ranging from Z=0.001 to Z=0.006 the
$M_V({\rm Bump})$ and the $M_V({\rm ZAHB})$ magnitudes changes
according to the following derivatives:
$\partial{M_V({\rm ZAHB})}/\partial{\rm [M/H]}\approx 0.093$
and $\partial{M_V({\rm Bump})}/\partial{\rm [M/H]}\approx 0.753$,
whereas for $0.006 \le Z \le 0.02$ they change according to 
$\approx0.289$ and $\approx1.251$ respectively.

The data plotted in the top panel of Figure 2 show clearly that the 
discrepancy of 0.4 magnitudes suggested by FP is completely removed 
over the entire metallicity range, and indeed the trend of empirical 
data is well reproduced by standard models.
This result is even more compelling if we take into account that the
previous comparisons (FP and CS) were hampered by the small number of
metal-poor clusters ([Fe/H] $< -1.5$) for which reliable estimates of
$V_{\rm Bump}$ were available. 
Figure 2 also extends the comparison to [Fe/H]$ = -0.2$; note how the 
observed values in the high metallicity range show the flattening predicted
by the theory, though with a distribution that is somehow flatter. 
A plausible change in the He content (CS) and/or the inclusion of 
element diffusion (CDS) can account for this effect only marginally. 
As a consequence, this result suggests that the metal-rich clusters 
could be younger than the bulk of our clusters (Salaris \& Weiss 1998; 
Rosenberg et al. 1999). 

In the cluster sample adopted in this investigation there are only
two clusters -NGC7078 and NGC5694- which are marginally in agreement
with the theoretical expectations.  For both clusters the uncertainty
on the location of the Bump is very small (Table 1). In the case of
NGC7078, the adopted mean $V_{\rm RR}$ magnitude is also in very good
agreement with the value obtained by Silberman \& Smith (1995).
Therefore for this cluster the discrepance could be due to an
underestimate of the cluster metallicity and/or of the $\alpha$
enhancement (see e.g. CG). An independent estimate of $V_{RR}$ for 
NGC5694 is not available, but any plausible assumption on its 
uncertainty can hardly remove such a discrepancy.

In order to account for the uncertainty on the metallicity scale 
the middle panel shows the same comparison as the top panel, but 
the observed points are plotted according to the Zinn \& West (1984, 
hereinafter ZW) metallicity scale. The data plotted in this panel show 
that the \vhbb values are systematically shifted toward lower metal 
contents when compared with theoretical observables. 
At present, both systematic and observational errors affecting 
the metallicity ranking of GGCs are still controversial issues (RHS; C96). 
This notwithstanding, the CG scale is more robust since it relies on 
recent high dispersion spectroscopic measurements and up-to-date 
atmosphere models. 
Even though current observational uncertainties affects the global
metallicity of individual clusters, data plotted in Figure 2 support 
the evidence that the ZW scale underestimates the cluster metallicity 
in the range $-1.7<$[Fe/H]$<-1.0$. In order to supply a quantitative 
estimate of the difference we performed a fit of the empirical
data ($-2.0<$[Fe/H]$<-0.5$) with predictions at 14 Gyr. The standard 
deviation is 0.05 mag for the CGBC scale and 0.17 mag for the ZW scale.
The latter value is almost a factor of two larger than the photometric
uncertainty. 


\section{Conclusions}

We have presented new homogeneous measurements of the
\vhbb values for a sample of 28 GGCs observed with HST,
and a detailed comparison with the theoretical models.  By relying on
homogeneous theoretical and observational frameworks and on the
metallicity scale suggested by CG and by CGBC we found
that, within current uncertainties, observables predicted by standard
H- and He-burning evolutionary models agree with the empirical
data.  This result is further strengthened by the fact that this
comparison was extended from metal-poor to metal-rich clusters 
($-2.1\le$ [Fe/H] $\le-0.2$).

New theoretical predictions for metal-rich clusters show a change of  
the slope of the \vhbb -[M/H] relation. This behavior is supported
by the tail of metal-rich clusters in our sample and does not depend 
on the adopted metallicity scale. Leading physical arguments on the  
dependence of \vhbb on input physics support the suggestion that 
metal-rich cluster could be younger than the bulk of clusters 
in our sample. 

By adopting the ZW metallicity scale we found that empirical data at 
low and intermediate metallicity are shifted toward lower metallicities 
when compared with theory. 
At the same time the agreement between theory and observations supports 
the use of a \vhbb - metallicity relation for constraining the 
cluster metallicity (Desidera 1999). However, we note that such a 
relation relies on the assumption that all GGCs are coeval within $\pm1$~Gyr 
(Stetson et al. 1999), and that the intrinsic accuracy is of the order 
of $0.15$~dex provided that the \vhbb values are measured with an 
accuracy of 0.10~mag.

We are deeply indebted to E. Carretta and R. Gratton for providing
us with the extension of the CG metallicity scale to [Fe/H]$=-0.3$  
in advance of publication. 


\pagebreak


\clearpage
\begin{deluxetable}{lclcc}
\tablewidth{0pt}
\tablecaption{Cluster Observables}
\tablehead{
\colhead{Object}& 
\colhead{$V_{\rm Bump}$}& 
\colhead{$V_{\rm RR}$}&  
\colhead{$V_{\rm RHB}$}&  
\colhead{$[Fe/H]_{CG}$}\nl
\colhead{(1)}&
\colhead{(2)}&
\colhead{(3)}&
\colhead{(4)}&
\colhead{(5)}}   
\startdata

104 ~~~47Tuc& 14.57$\pm 0.02$ &\nodata& 14.16 & -0.70$^a$\nl
362	     & 15.47$\pm 0.02$ & 15.33 &\nodata& -1.15$^a$\nl
1851         & 16.16$\pm 0.02$ & 16.05 &\nodata& -1.14$^b$\nl
1904 ~~M79  & 16.00$\pm 0.04$ & 16.21 &\nodata& -1.37$^a$\nl
2808         & 16.31$\pm 0.03$ & 16.25 &\nodata& -1.24$^b$\nl
5634         & 17.77$\pm 0.03$ & 17.95 &\nodata& -1.66$^b$\nl
5694         & 18.49$\pm 0.04$ & 18.64 &\nodata& -1.70$^b$\nl
5824         & 18.10$\pm 0.03$ & 18.46 &\nodata& -1.69$^b$\nl
5927         & 17.37$\pm 0.04$ &\nodata& 16.83 & -0.31$^b$\nl
6093 ~~M80  & 16.12$\pm 0.03$ & 16.36 &\nodata& -1.48$^b$\nl
6139         & 18.30$\pm 0.03$ & 18.40 &\nodata& -1.50$^b$\nl
6205 ~~M13  & 14.70$\pm 0.04$ & 14.99 &\nodata& -1.39$^a$\nl
6235         & 17.24$\pm 0.05$ & 17.31 &\nodata& -1.27$^b$\nl
6273 ~~M19  & 16.77$\pm 0.04$ & 16.91 &\nodata& -1.53$^b$\nl
6284         & 17.36$\pm 0.05$ & 17.25 &\nodata& -1.20$^b$\nl
6287         & 16.60$\pm 0.05$ & 17.21 &\nodata& -1.88$^b$\nl
6293         & 16.04$\pm 0.05$ & 16.47 &\nodata& -1.76$^b$\nl
6342         & 18.00$\pm 0.05$ &\nodata& 17.56 & -0.57$^b$\nl
6356         & 18.53$\pm 0.03$ &\nodata& 18.05 & -0.44$^b$\nl
6362         & 15.60$\pm 0.02$ & 15.35 &\nodata& -0.96$^a$\nl
6388         & 17.69$\pm 0.04$ &\nodata& 17.31 & -0.53$^b$\nl
6441         & 18.46$\pm 0.04$ &\nodata& 17.99 & -0.46$^b$\nl
6522         & 17.01$\pm 0.04$ & 17.06 &\nodata& -1.38$^b$\nl
6624         & 16.68$\pm 0.02$ &\nodata& 16.19 & -0.36$^b$\nl
6652         & 16.44$\pm 0.02$ &\nodata& 16.11 & -0.86$^b$\nl
6934         & 16.85$\pm 0.03$ & 17.01 &\nodata& -1.40$^b$\nl
6981 ~~M72  & 17.13$\pm 0.04$ & 17.26 &\nodata& -1.40$^b$\nl
7078 ~~M15  & 15.41$\pm 0.04$ & 15.86 &\nodata& -2.12$^a$\nl
\enddata
\tablenotetext{a}{CG metallicities from high resolution spectroscopy.}
\tablenotetext{b}{Cluster metallicities based on CGBC scale.}  
\end{deluxetable}
\clearpage
\begin{figure}
\plotone{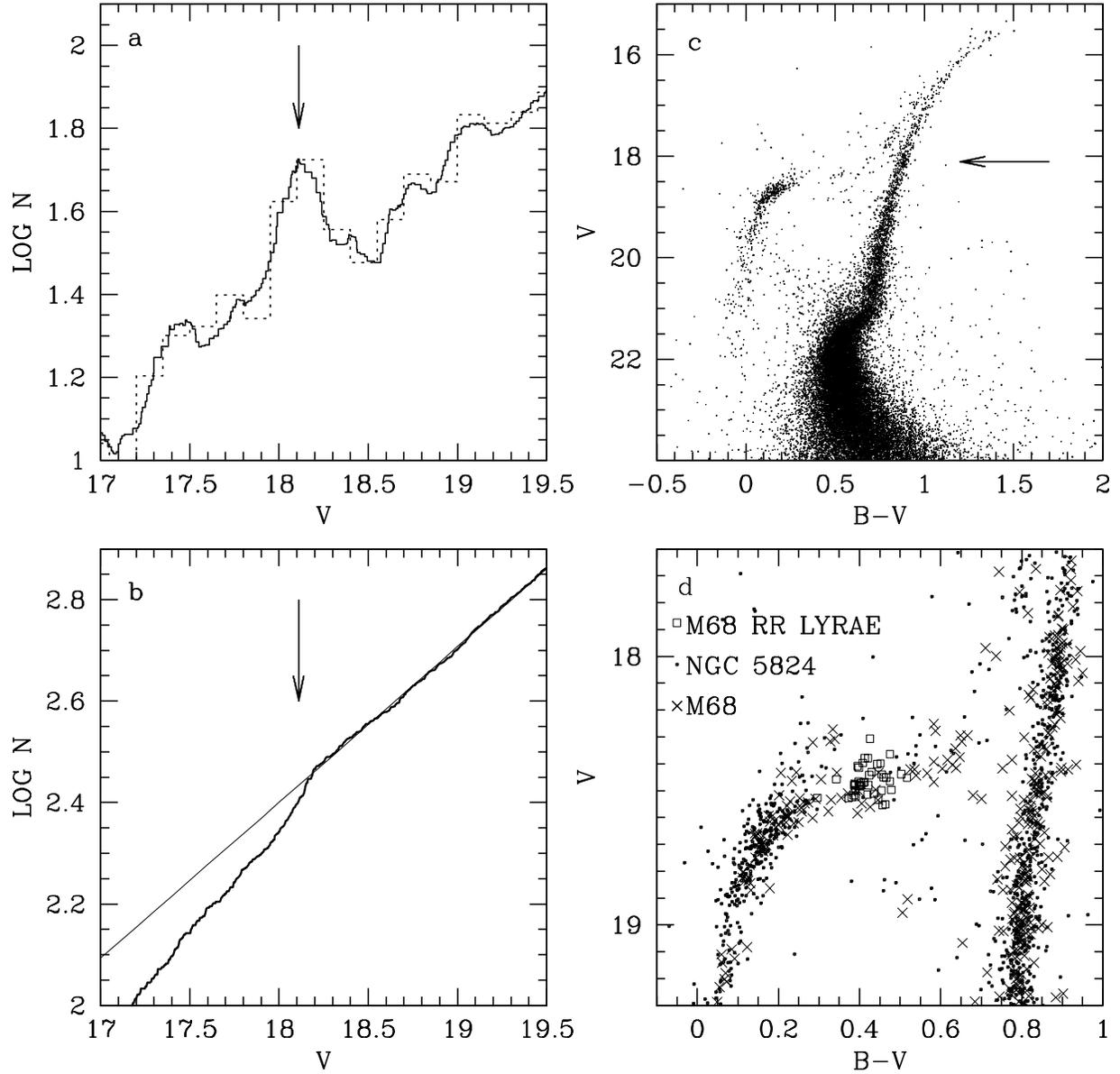}
\caption{Example of the empirical measure of the Bump and ZAHB location.
The arrows indicate the adopted Bump magnitude.
{\em Panel a}: Differential LF; {\em Panel b}: cumulative LF; 
{\em Panel c}: position of the Bump in the CMD; 
{\em Panel d}: match between the HBs of NGC5824 and of the template 
cluster M68. \label{fig1}}
\end{figure}
\clearpage
\begin{figure}
\plotone{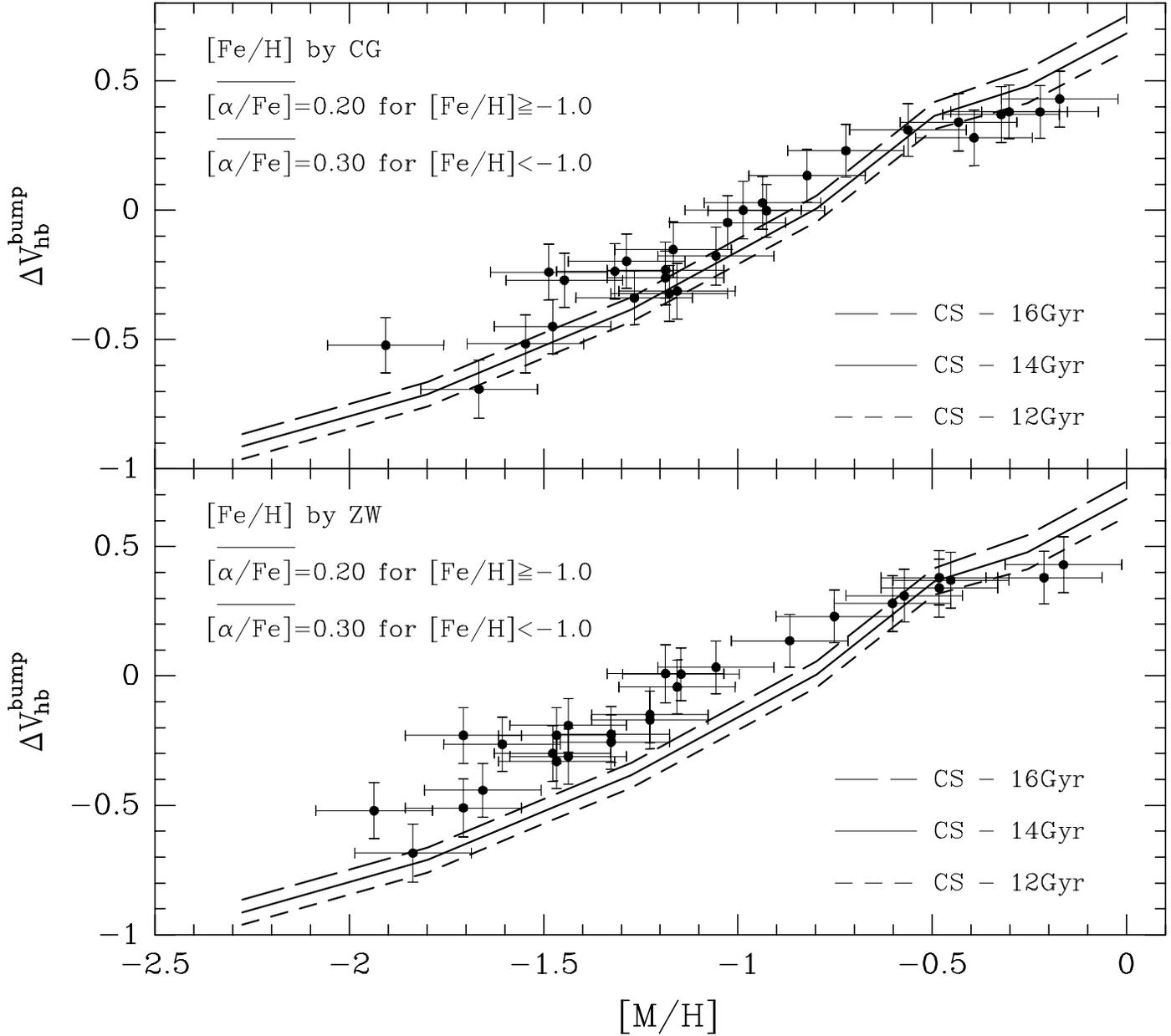}
\caption{Comparison between the theoretical and empirical values of 
\vhbb as a function of the global metallicity. {\em Top panel}: the
empirical data plotted according to the CG metallicity scale (see
text for more details). {\em Bottom panel}: Same as the top panel but with the 
empirical data were plotted according to the ZW metallicity scale.\label{fig2}}
\end{figure}
\end{document}